\documentclass[aps,prl,reprint,superscriptaddress]{revtex4-1}
\usepackage{color}
\usepackage{graphicx}
\usepackage{natbib}
\usepackage{url}
\usepackage{textcase}
\usepackage{bm}
\usepackage{amsmath} 
\usepackage{ulem} 

\begin{document}

\title{Critical Slowing Down of the Charge Carrier Dynamics\\ at the 
Mott Metal-Insulator Transition}%

\author{Benedikt Hartmann}
\affiliation{Institute of Physics and SFB/TR\,49, Goethe-University Frankfurt, 60438 Frankfurt (M), Germany}
\author{David Zielke}
\affiliation{Institute of Physics and SFB/TR\,49, Goethe-University Frankfurt, 60438 Frankfurt (M), Germany}
\author{Jana Polzin}
\affiliation{Institute of Physics and SFB/TR\,49, Goethe-University Frankfurt, 60438 Frankfurt (M), Germany}
\author{Takahiko Sasaki}
\affiliation{Institute for Materials Research, Tohoku University, Sendai, 980-8577, Japan}
\author{Jens M\"uller}
\email[Email: ]{j.mueller@physik.uni-frankfurt.de}
\affiliation{Institute of Physics and SFB/TR\,49, Goethe-University Frankfurt, 60438 Frankfurt (M), Germany}

\date{\today}

\begin{abstract}
We report on the dramatic slowing down of the charge carrier dynamics in a quasi-two-dimensional organic conductor, which can be reversibly tuned through the 
Mott metal-insulator transition (MIT). 
At the finite-temperature critical endpoint we observe a divergent increase of the resistance fluctuations accompanied by a drastic shift of spectral weight to low frequencies, demonstrating the critical slowing down of the order parameter (doublon density) fluctuations. 
The slow dynamics is accompanied by non-Gaussian fluctuations, indicative of correlated charge carrier dynamics. A possible explanation is a glassy freezing of the electronic system as a precursor of the Mott MIT.  
\end{abstract}

\maketitle

\begin{figure}[h]
\includegraphics[width=\columnwidth]{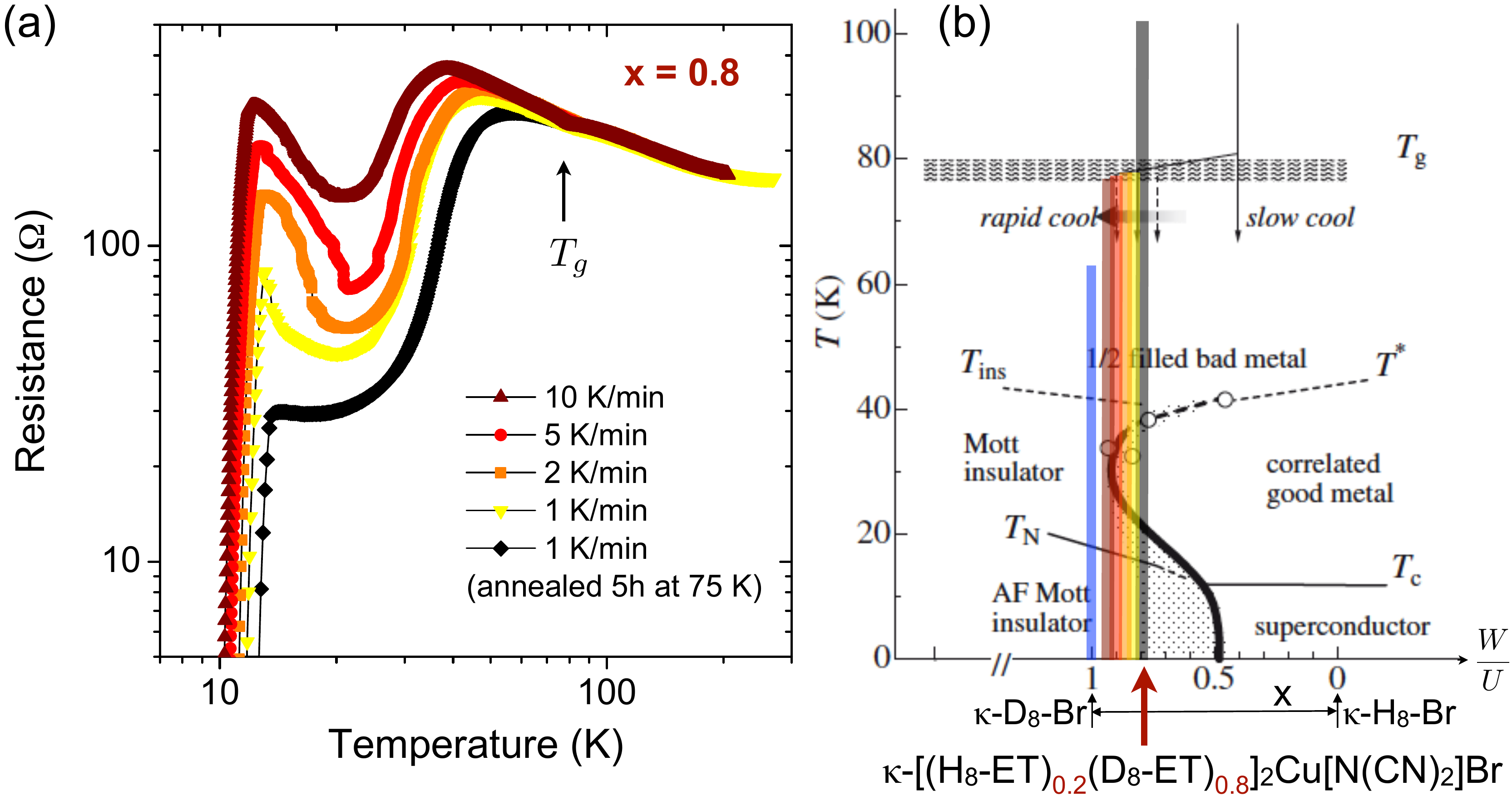} 
\caption{\label{Fig1}(Color online) (a) Resistance measurements of $\kappa$-D$_8$/H$_8$-Br for various cooling rates $q$. 
(b) Arrow indicates the position of the slowly-cooled, pristine sample in the generic phase diagram. Open circles indicate values taken from the literature of the second-order critical end point of the first-order Mott MIT (thick solid curve), 
see \cite{Sasaki2005} and references therein. 
The change of the sample position with $q$ is schematically indicated by the colored lines. Blue represents fully deuterated $\kappa$-D$_8$-Br \cite{Brandenburg2012}.}
\end{figure}

The Mott metal-insulator transition (MIT), where a charge gap opens due to electron-electron interactions is a key phenomenon in modern condensed-matter physics \cite{Imada1998},  
for which the understanding of various fundamental aspects remains challenging, both theoretically and experimentally \cite{Georges2004}. Among them are the nature of the anomalous metallic state and (nano-scale) phase separation in the vicinity of the Mott transition, the understanding of the combined effects of electron-electron interactions and disorder, and the question of universality and critical behavior. 
In recent years, organic charge-transfer salts have proven an outstanding class of materials with model character for studying these aspects of the Mott MIT \cite{Georges2004,Toyota2007,Kanoda2008,Powell2011}. These materials $\kappa$-(BEDT-TTF)$_2$X, composed of the organic donor molecule BEDT-TTF (bis-ethylenedithio-tetrathiafulvalene representing C$_6$S$_8$[C$_2$H$_4$]$_2$, abbreviated as ET) and a polymeric anion X, are layered systems with a quasi-two-dimensional electronic structure and belong to the very few examples, where the Mott MIT is purely driven by controlling the bandwidth without symmetry breaking \cite{Kagawa2009}. The bandwidth $W$ 
can be covontrolled either continuously by applying hydrostatic pressure or in discrete steps at ambient pressure by varying the anion X, which mimics changes in the ratio $W/U$, where $U$ is the effective on-site Coulomb repulsion \cite{Kanoda1997}. 
Likewise, replacing the eight H-atoms of the ET molecules' C$_2$H$_4$ ethylene endgroups (EEG) in metallic $\kappa$-(ET)$_2$Cu[N(CN)$_2$]Br ($\kappa$-H$_8$-Br) by D-atoms ($\kappa$-D$_8$-Br) results in a chemically-induced shift from the metallic towards the Mott insulating state \cite{Kawamoto1997}. In the $T$-$p$ or $T$-X phase diagram the MIT is represented by an S-shaped, first-order transition line which terminates in a second-order critical point $(p_{cr.}/T_{cr.})$, see Fig.\,\ref{Fig1}. 
Recently, the critical behavior of the charge, spin and lattice degrees of freedom at $(p_{cr.}/T_{cr.})$ has been subject of intense research efforts \cite{Kagawa2005,Papanikolaou2008,Kagawa2009,Bartosch2010,Zacharias2012,Sordi2012,Furukawa2015,Abdel-Jawad2015}. 
Despite numerous experimental and theoretical approaches, however, no consensus has been reached on the nature of the Mott criticality in $\kappa$-(ET)$_2$X and the underlying universality class. Furthermore, an investigation of the critical fluctuations and the  dynamic properties of the charge carriers at the critical point, in particular at low frequencies, is still lacking. From general arguments, a critical slowing down of the fluctuations is predicted for a second-order phase transition \cite{Mazenko2006} but has not yet been observed at the Mott critical endpoint \cite{Kagawa2009}.\\
In this letter, we present a systematic study of the dynamical properties of the electrons at the Mott transition and critical endpoint using fluctuation spectroscopy, a technique, where the dynamics of charge carriers is studied without injecting additional electrons into the system. For our study, we have chosen the material $\kappa$-[(H$_8$-ET)$_{0.2}$(D$_8$-ET)$_{0.8}$]$_2$Cu[N(CN)$_2$]Br ($\kappa$-D$_8$/H$_8$-Br in short), where the partial substitution of the ET molecules with their deuterated analogues places the sample 
on the high-pressure (metallic) side of the phase diagram very close to the critical pressure $p_{cr.}$ of the MIT \cite{Sasaki2005}. 
Upon fine-tuning the material through the transition we find an electronic-correlation-induced enhancement of the low-frequency resistance fluctuations, which strongly increase and nearly diverge when approaching the critical point, accompanied by a dramatic slowing down of the fluctuation dynamics. At the temperatures, where the fluctuations are largest, non-Gaussian fluctuations are observed. 

$\kappa$-[(H$_8$-ET)$_{0.2}$(D$_8$-ET)$_{0.8}$]$_2$Cu[N(CN)$_2$]Br   
has been grown as single crystals by electrochemical crystallization \cite{Yoneyama2004}. Low-frequency fluctuation spectroscopy measurements have been performed in a five-terminal setup using a bridge-circuit ac technique \cite{Scofield1987} described in detail in \cite{Mueller2011}. 
Resistance $R$ and resistance noise power spectral density (PSD) $S_R(f) = 2 \langle |\delta \tilde{V}(f)|^2 \rangle/I^2$ \cite{comment_v-noise} have been measured perpendicular to the conducting layers, where 
$I$ is the applied current. For all measurements discussed here, we have observed excess noise of general $S_R \propto 1/f^\alpha$-type (see \cite{Mueller2009,Brandenburg2012} for typical spectra), characterizing the intrinsic resistance (conductance) fluctuations of the sample. 
%

Figure\,\ref{Fig1}(a) shows the resistance of $\kappa$-D$_8$/H$_8$-Br for various rates of cooling $q = {\rm d}T/{\rm d}t$ through the temperature $T_g \approx 75 - 80$\,K of the glass-like EEG ordering \cite{Mueller2002,Hartmann2014}, which results in a frozen non-equilibrium occupation of the two possible EEG conformations \cite{comment_glass-disorder}. The transformation from metallic to insulating behavior with increasing $q(T_g)$, see Fig.\,\ref{Fig1}, has been widely used in $\kappa$-Br compounds with different type and degree of deuteration of the ET molecules \cite{Taniguchi1999,Taniguchi2003,Yoneyama2004,Sasaki2005}. The effect originates in an anisotropic change of the in-plane lattice parameters at $T_{g}$, 
which changes the relevant transfer integrals such that more rapid cooling leads to slightly smaller bandwidth (and $W/U$ ratio) \cite{Sasaki2005,Hartmann2014}. 
In Fig.\,\ref{Fig1}(a), the sample 
which has been annealed for 5\,h in the vicinity of $T_g$, shows essentially metallic behavior. 
In contrast, the resistance curves taken during continuous cool-down measurements with increasing $q(T_g)$ 
reveal the typical re-entrant behavior, where the S-shaped first-order Mott transition line is crossed twice. In between the slowest and fastest cooling-rate of the experiment, the sample has crossed the second-order critical endpoint $(p_{cr.},T_{cr.})$ of the Mott transition. All cool-down measurements shown here reveal a superconducting transition at low temperatures, which 
is of inhomogeneous, percolating type, due to the coexistence of metallic/superconducting and insulating/antiferromagnetic phases \cite{Yoneyama2004,Kagawa2004,Mueller2009b}. 
We note that this fine-tuning of the electronic bandwidth by controlling the cooling rate is reversible, {\it i.e.}\ upon warming above $T_g$ the frozen EEG glass melts and the lattice relaxes. 

\begin{figure*}[htb]
\includegraphics[width=\textwidth]{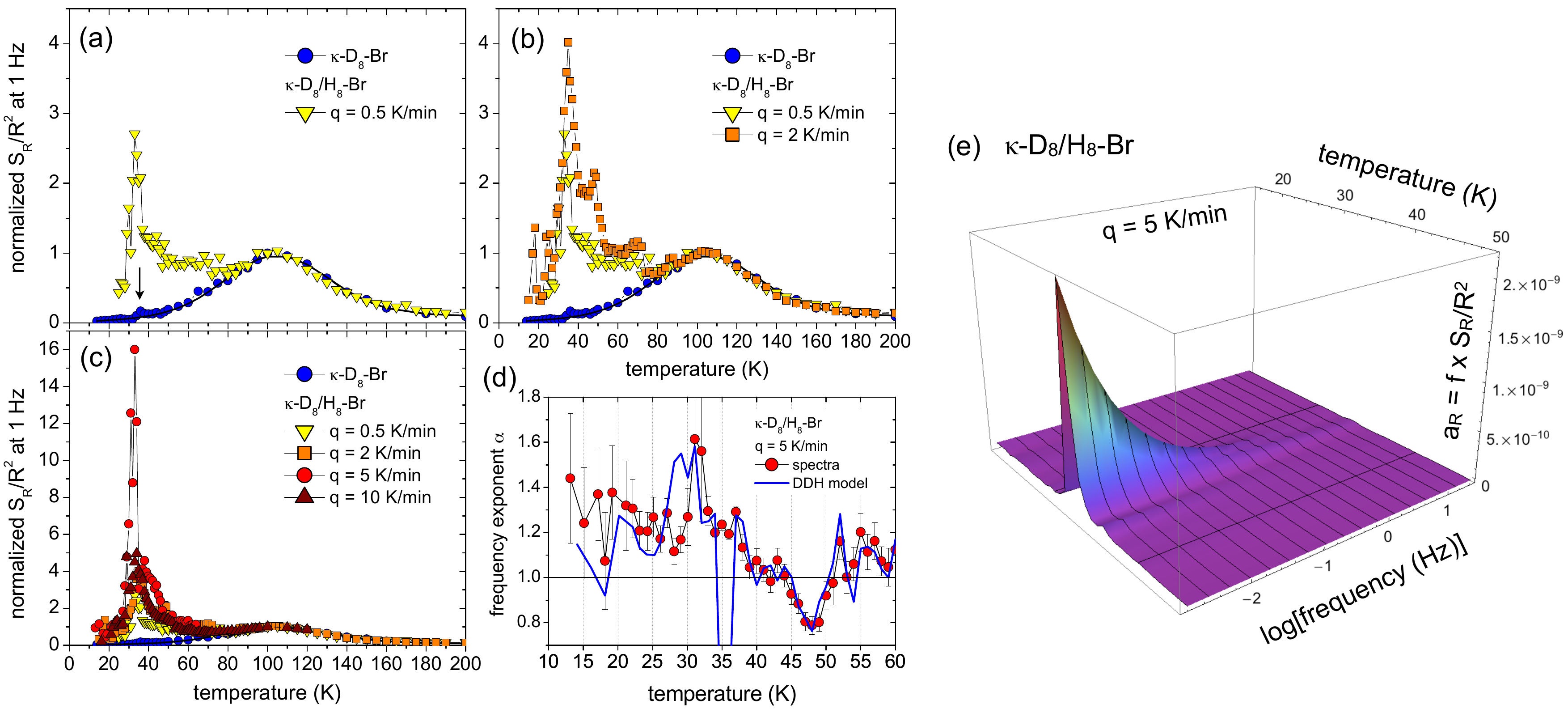} 
\caption{\label{Fig2}(Color online) (a) - (c) Resistance noise (PSD) taken at $f = 1$\,Hz of $\kappa$-D$_8$/H$_8$-Br compared to $\kappa$-D$_8$-Br (slowly cooled, same data as in \cite{Brandenburg2012}) normalized to the value of the local maximum around 100\,K 
for increasing cooling rates (a) $q = 0.5$, (b) $2$, (c) $5$ and $10$\,K/min. Note the rescaling of the ordinate in (c). (d) Frequency exponent $\alpha$ {\it vs.}\ $T$ of the $1/f^\alpha$-type noise spectra for $q = 5$\,K/min. Line is a description with a model of non-exponetial kinetics (DDH model) as described in the text. (e) Relative noise level $a_R \equiv f \times S_R/R^2$ {\it vs.}\ $f$ {\it vs.}\ $T$ for the same cooling rate.}
\end{figure*}
After the continuous cool-down runs with different rates $q$, noise measurements have been performed in each case upon warming the sample in discrete temperature steps. Each data point in Fig.\,\ref{Fig2} represents one $1/f^\alpha$-type spectrum in the frequency range between 1\,mHz and 100\,Hz, where the magnitude of the fluctuations, $S_R/R^2(f = 1\,{\rm Hz})$, and the frequency exponent, $\alpha \equiv \partial \ln{S_R}/\partial \ln{f}$, have been evaluated. Figure\,\ref{Fig2}(a) - (c) shows the resistance noise PSD for $\kappa$-D$_8$/H$_8$-Br in comparison to the fully deuterated compound $\kappa$-D$_8$-Br (data from \cite{Brandenburg2012}), which is located slightly on the insulating side of Mott transition, see Fig.\,\ref{Fig1}(b). The data are normalized to the broad maximum in $S_R/R^2$ at around 100\,K, which is due to the coupling of the charge carriers to the fluctuating EEG of the ET molecules \cite{Mueller2009,Brandenburg2012,Hartmann2014}, which undergo a glass-like ordering transition at $T_g \sim 75 - 80$\,K as described above. Since this feature is of structural origin, it is 
independent of the samples' position in the generalized phase diagram. Figure\,\ref{Fig2}(a) compares the slowly-cooled fully and partially deuterated compounds $\kappa$-D$_8$-Br and $\kappa$-D$_8$/H$_8$-Br, respectively, showing essentially the same behavior at high temperatures $T \gtrsim T_g$. For $\kappa$-D$_8$-Br, the arrow indicates a sharp  peak in the noise at $T = 36$\,K (however, small in comparison to the broad maximum at 100\,K), which grows larger at frequencies smaller than 1\,Hz (see Figs.\,1 and 3 in \cite{Brandenburg2012}). This feature, which is absent in the fully-hydrogenated metallic system $\kappa$-H$_8$-Br (not shown), has been interpreted as a sudden slowing down of the charge carrier dynamics due to the vicinity of the Mott critical endpoint \cite{Brandenburg2012}.
In contrast to $\kappa$-D$_8$-Br, we observe a larger overall noise level for $\kappa$-D$_8$/H$_8$-Br and even a slight increase below $T_g \sim 75 - 80$\,K, {\it i.e.}\ where the conformational vibrations of the EEG are frozen out. This can be understood considering the sample's position in the phase diagram closer to the Mott insulating phase resulting in a correlation-induced increase of the low-frequency fluctuations, since the charge carriers already tend to localize, an effect, which possibly is enhanced by weak static disorder \cite{Diehl2015}. Likewise, electronic phase separation in this region may enhance the low-frequency fluctuations \cite{Mueller2009b,Rommel2013}. 
Most striking, however, is the sharp and -- compared to $\kappa$-D$_8$-Br -- strongly enhanced noise peak at $T \sim 33$\,K for the $\kappa$-D$_8$/H$_8$-Br sample cooled with $q = 0.5$\,K/min. 

These large low-frequency fluctuations become enhanced even more dramatically for increasing cooling rates of $q = 2$\,K/min and 5\,K/min shown in Figs.\,\ref{Fig2}(b) and (c), respectively \cite{comment_finestructure}. Note the rescaled ordinate in Figs.\,\ref{Fig2}(c) demonstrating the very large noise level observed for the data taken at 5\,K/min, {\it i.e.}\ this particular position in the phase diagram, while for $q = 10$\,K/min the noise peak decreases again. This observation (i) rules out a mere effect of increasing lattice disorder \cite{comment_glass-disorder}, and (ii) suggests that the curve $q = 5$\,K/min is the one closest to the critical endpoint of the Mott transition, with the highest noise level at $T_{cr.} = 33$\,K. 
The frequency exponent $\alpha(T)$ for this curve, shown in Fig.\,\ref{Fig2}(d), increases upon approaching $T_{cr.}$ from values of 0.8 at $T = 48$\,K to $\alpha = 1.6$ at $31$\,K, corresponding to a strong shift of spectral weight to low frequencies. In general, large values of $\alpha > 1.4$ are typical for a system far from equilibrium, as we will discuss below. At the lowest frequency of our experiment of $f = 1$\,mHz, the noise level increases by more than three orders of magnitude upon approaching the critical point 
from above. In the study of dynamical critical phenomena the existence of such large times is known as critical slowing down of the order parameter relaxation rate 
\cite{Mazenko2006}, where the length scale $\xi$, which measures the correlations of the order parameter becomes infinite upon approaching a second-order phase transition. The time evolution of the order parameter becomes very slow since the correlated regions in the sample involve very many degrees of freedom. Therefore, the kinetics of the order parameter near a critical point causes a slow dynamics of a macroscopic system, in our case manifested in the diverging resistance/conductance fluctuations. The expected singular nature of the critical point as probed by order-parameter fluctuations is highlighted in Fig.\,\ref{Fig2}(e), where we plot the relative noise level $a_R(f,T) \equiv f \times S_R/R^2$, a dimensionless number characterizing the strength of the fluctuations. 
Our results 
demonstrate the expected divergence of the doublon (two electrons sitting on the same atomic orbitals) susceptibility 
and the critical slowing down of doublon density fluctuations \cite{Kotliar2000,Imada2005,Kagawa2009}.

\begin{figure}[h]
\includegraphics[width=0.7\columnwidth]{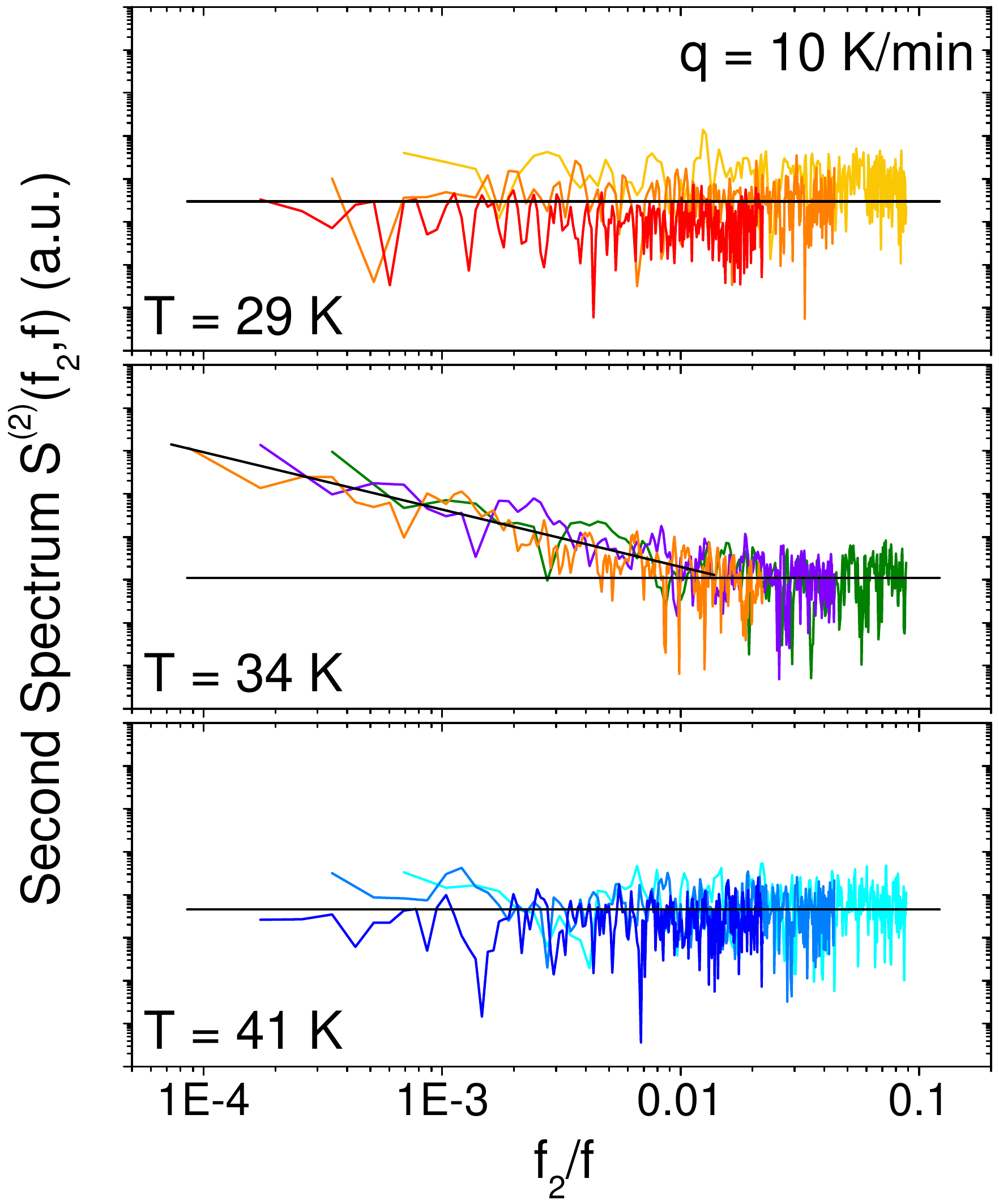} 
\caption{\label{Fig3}(Color online) Second spectra $S^{(2)}(f_2,f)$ {\it vs.} $f_2/f$ of $\kappa$-D$_8$/H$_8$-Br for $q = 10$\,K/min at three different temperatures. Horizontal lines are guides for the eyes. At $T = 34$\,K, where the first noise spectra $S_R/R^2(f)$ peaks at $f = 1$\,Hz, see Fig.\,\ref{Fig2}(c), the data at low frequencies can be fitted to $S^{(2)} \propto 1/f_2^{\beta}$ yielding $\beta = 1.35$ (black line). At all temperatures the spectra measured at different frequency octaves are scale invariant, {\it i.e.}\ do not depend on the scale of $f$.} 
\end{figure}

Besides the strongly increasing amplitude of the low-frequency fluctuations at the critical point, another important aspect related to the dynamic criticality can be deduced from our data. As we have shown in our earlier studies \cite{Mueller2009,Brandenburg2012}, the $1/f$-type fluctuation properties of the organic charge-transfer salts $\kappa$-(ET)$_2$ can be very well described by a model of non-exponential kinetics \cite{Kogan1996}, where the $1/f^\alpha$-noise originates in the superposition of a large number of {\it independent}, {\it i.e.}\ Gaussian, so-called 'fluctuators'. Within this model, first introduced by Dutta, Dimon and Horn (DDH) \cite{Dutta1979}, a distribution of activation energies $D(E)$ of the individual fluctuations determines 
the observed frequency and temperature dependence $S_R(f,T)$. The assumptions of the model can be checked for consistency by comparing
\begin{equation}
\alpha_{\rm DDH}(T) = 1 - \frac{1}{\ln 2 \pi f \tau_0} \biggl(\frac{\partial \ln S(f,T)}{\partial \ln T} - 1\biggr).
\label{eq-alpha}
\end{equation}
with the measured data $\alpha(T)$ \cite{comment_DDH}. The blue line in Fig.\,\ref{Fig2}(d) shows $\alpha_{\rm DDH}(T)$, which describes the measured data very well, except for an obvious deviation just above the critical point at $T_{cr.}$, where the noise level peaks. We note that such a deviation, which we also see for the $q = 10$\,K/min curve, is not observed for $\kappa$-D$_8$-Br \cite{Brandenburg2012} and for the smaller cooling rates of the present sample $\kappa$-D$_8$/H$_8$-Br (not shown). An obvious interpretation is that the assumption of independent two-level fluctuators is simply not valid in a narrow temperature range around $T_{cr.}$, which is a first hint to non-Gaussian fluctuations near the critical point.\\ 
In order to further explore this matter, we introduce the so-called second spectrum $S^{(2)}(f_2,f,T)$, which is the power spectrum of the fluctuations of $S_R(f,T)$ with time, {\it i.e.}, the Fourier transform of the autocorrelation function of the time series of $S_R(f)$ \cite{Kogan1996}, introducing an additional frequency $f_2$ related to the time over which $S_R(f)$ fluctuates \cite{comment_2nd-spectra}. $S^{(2)}(f_2,f)$ probes 
a fourth-order noise statistics and therefore deviations from Gaussian behavior: it is independent of the frequency $f_2$ ("white") 
if the fluctuations are uncorrelated, {\it i.e.}\ caused by independent two-level systems. In contrast, $S^{(2)} \propto 1/f_2^\beta$ for correlated (interacting) fluctuators \cite{Weissman1982,Weissman1993}. As shown in Fig.\,\ref{Fig3} for $\kappa$-D$_8$/H$_8$-Br cooled-down with $q = 10$\,K/min, {\it i.e.}, even slightly beyond the critical point, the second spectrum is 'white' above and below the temperature of the noise peak in the first spectrum. At $T = 34$\,K, where the noise peaks, however, $S^{(2)}$ is frequency-dependent with a large exponent $\beta = 1.35$, see black line in Fig.\,\ref{Fig3}. Clearly, close to the critical point, where the fluctuation dynamics is slow, this spectral wandering indicates correlated fluctuations, which rules out independent two-level systems as the source of the noise. 

In general, non-Gaussian fluctuations caused by slow modes corresponding to large scales are expected near a critical point of a second-order phase transition and have been observed, {\it e.g.}, in liquid crystals \cite{Joubaud2008}.
However, since non-ergodic dynamics is found in a wider range of bandwidths around the critical region, {\it i.e.}, for $q = 5$, 10\,K/min and faster cooling rates, it is interesting to compare our findings to noise measurements of other systems showing metal-insulator transitions. For a 2D MIT in Si inversion layers, a similar dramatic slowing down of the electron dynamics is observed \cite{Bogdanovich2002}, which is interpreted as a glassy freezing of the electron system, in agreement with the onset of non-Gaussian fluctuations already in the metallic phase. 
Similar observations for the 3D Anderson-Mott MIT in P-doped Si are likewise interpreted as a glassy freezing of the electrons preceding their localization \cite{Kar2003},  and the intriguing possibility that such correlated dynamics may be a universal feature of MITs, irrespective of the systems' dimensionality, is raised. We note that in these systems the noise diverges only for $T \rightarrow 0$, whereas for the present system a drastic increase and slowing down of the charge-carrying fluctuations is observed for the first time at a finite-temperature critical endpoint. A glassy freezing of the charge carriers near the critical point may be a possible scenario also for the present strongly-correlated materials \cite{Schmalian2000,Dobro2003}. Here, a residual structural disorder potential and/or the inherent frustration of the triangular lattice geometry of the ET molecules forming dimers could result in a large number of metastable states, which the correlated electrons have to overcome. General models for such glassy dynamics applied to spin glasses are often invoked also for the charge degrees of freedom \cite{Jaro2002}. In a model of interacting droplets/clusters $S^{(2)}(f_2,f)$ should be a decreasing function of $f$ for constant $f_2/f$ \cite{Weissman1982,Jaro2002}. The observed scale invariance shown in Fig.\,\ref{Fig3} (note the different colors for each frequency octave), however, points to a hierarchical picture, where the system wanders collectively between metastable states related by a kinetic hierarchy \cite{Weissman1982}. The large exponent of $\beta = 1.35$ indicates that only few such states are visited during the time of measurement. From the theoretical point of view, a Coulomb glass behavior in the vicinity of the Mott transition remains controversial. For example, a self-generated glass transition caused by the frustrated nature of the interactions (and not related to the presence of quenched disorder) is predicted in doped Mott insulators \cite{Schmalian2000}. In \cite{Dobro2003}, however, the metallic glassy phase is suggested to be suppressed for Mott localization and to become stabilized only for increasing disorder, in apparent agreement with experimental results on Si inversion layers with different degrees of disorder. 

Our method applied to the present highly-tunable molecular conductors opens the door to systematically explore the dynamic critical properties of strongly-correlated electrons in the presence of disorder and frustration.
Whether a universal electron glass state emerges in the vicinity of the finite-temperature critical endpoint of the Mott transition in $\kappa$-(ET)$_2$X or non-ergodic dynamics and the relevance of higher statistical moments is an inherent characteristics of the proximity to the critical point of the second-order phase transition will be seen in future systematic studies of fluctuation spectroscopy for different bandwidths and disorder levels.
Furthermore, in view of the large anomalies in the sound velocity at the critical point, which is consistent with a diverging compressibility of the electronic degrees of freedom \cite{Fournier2003}, and in the lattice response at the critical point \cite{deSouza2007}, the coupling of the diverging low-frequency fluctuations to the lattice degrees of freedom are an important field of investigations in the future.\\

We acknowledge financial support from the Deutsche Forschungsgemeinschaft (DFG) within the collaborative research center SFB/TR 49 and acknowledge funding from Grant-in-Aid for Scientific Research (B) from JSPS KAKENHI, Japan (Grant No.\ 25287080). 




\begin{thebibliography}{99}
\bibitem{comment_finestructure} The observed finestructure (smaller peaks in the noise PSD at different temperatures) is reproducible and is likely due to sample inhomogeneities and electronic phase separation caused by the first-order nature of the Mott transition.
\bibitem{comment_2nd-spectra} The 'second noise spectra' $S^{(2)}(f_2,f)$ have been calculated by dividing the first spectra $S(f)$ into octaves and digitalizing the data by summing up the points (of the first spectra) within the octaves. A spectral analysis of these data is then performed for each octave separately. This introduces a new frequency $f_2 = 1/T_1$, with $T_1$ the measuring time for the first spectra.
\end{thebibliography}
\end{document}